\documentclass[twocolumn,apjl]{emulateapj} 
\usepackage{times}
\usepackage{natbib}
\usepackage{rotating}

\newcommand{\nii}{{[N\,{\sc ii}]}}

\newcommand{\hii}{H\,{\sc ii}\rm}

\newcommand{\ha}{{H$\alpha$}}

\newcommand{\lam}{$\,\lambda$}
\newcommand{\llam}{$\,\lambda\lambda$}

\begin{document}
\title{Systematics in Metallicity Gradient Measurements I : Angular Resolution, Signal-to-Noise and Annuli Binning}
\author{T.-T. Yuan\altaffilmark{1,2}, L. J. Kewley\altaffilmark{1,2}, \& J. Rich\altaffilmark{3}}
\altaffiltext{1}{Institute for Astronomy, University of Hawaii, 2680 Woodlawn Drive, Honolulu, HI 96822}
\altaffiltext{2}{Research School of Astronomy and Astrophysics, The Australian National University, Cotter Road, Weston Creek, ACT 2611}
\altaffiltext{3}{Carnegie Observatories, 813 Santa Barbara Street, Pasadena, CA 91101, U.S.A.}

\begin{abstract}
With the rapid progress in metallicity gradient studies at high-redshift, it is imperative that we thoroughly understand the systematics in these measurements.
This work investigates how the   \nii\,/\ha\ ratio based metallicity gradients  change with angular resolution, signal-to-noise (S/N), and annular binning parameters. 
Two approaches are used:  1. We downgrade the high angular resolution integral-field  data of  a gravitationally lensed galaxy and re-derive the metallicity
gradients at different angular resolution;  2. We simulate high-redshift integral field spectroscopy (IFS) observations under different angular resolution and S/N conditions using a local galaxy with a known gradient.   We find that the measured metallicity gradient changes systematically with angular resolution and annular binning.  Seeing-limited observations
produce significantly flatter gradients than higher angular resolution observations.  There is a critical angular resolution limit beyond  which the measured metallicity  gradient 
is substantially different to the intrinsic gradient.   This critical angular resolution  depends on the intrinsic gradient of the galaxy and is $\leq0.02^{\prime\prime}$ for our simulated galaxy.   
We show that seeing-limited high-redshift metallicity gradients are likely to be strongly affected by resolution-driven gradient flattening.
Annular binning with a small number of annuli produces a more flattened gradient than the intrinsic gradient due to weak line smearing. 
For 3-annuli bins, a minimum S/N of $\sim 5$ on the \nii\ line is required for the faintest annulus to constrain the gradients with meaningful errors.  
\end{abstract}


\section{Introduction}
The chemical abundance distribution within galaxies at local and high redshift offers unique insights
into the formation and evolution of galaxies.   Indeed,  the 
first simple formation models of our Milky Way  were built upon observational knowledge of the 
Galactic chemical distribution and stellar dynamics \citep{Eggen62,Searle78}.

For local galaxies, the existence of radial metallicity gradients has been established in most spiral galaxies using
abundances in \hii\ regions \citep{Pagel81,Edmunds84,Evans86,Garnett87,Vilchez88,Shields90,VilaCostas92,Zaritsky94,Dutil99}.  
The radial metallicity gradients are negative, with a higher metallicity towards the galactic center.  On the other hand, mergers and 
 barred galaxies have shallower or flattened gradients  due to  interaction-induced gas flows
 or bar induced gas inflows \citep{Kewley10,Rupke10a,Rupke10b}.

In the classic galactic chemical evolution models,   the radial  metallicity gradients formed during  inside-out  galaxy mass assembly.
In this scenario, gas infall timescales increase with galactocentric  distance \citep{Prantzos00,Chiappini01,Molla05,Magrini07,FuJ09}.  However, even among inside-out formation models, no consensus has been reached on the direction and magnitude  of the cosmic time evolution of the metallicity gradient.  

The  time evolution of the metallicity gradient in disk galaxies provides tight constraints on the historical events that mark
the galactic disc structure evolution.  The cosmic metallicity gradient evolution is directly linked with  the size-growth, external gas accretion,  galactic-scale outflows, and 
any internal transportation of the star-forming gas in galaxies.  Early analytical models such as \citet{Edmunds95} have provided perceptive predictions on how the gas flows can 
change the origin and evolution of the metallicity gradient.  Current $\Lambda$CDM cosmological hydrodynamic simulations now  incorporate the prescriptions for galactic chemical evolution models and  initial predictions  for the metallicity gradient evolution with redshift are beginning to emerge \citep[e.g.,][]{Pilkington12b}.  
To  guide and compare with  these models and simulations, it is of crucial importance to establish an observational baseline for the time evolution of metallicity gradients in galaxies.

Observational constraints on the time evolution of the metallicity gradient have been scarce and are traditionally based on indirect methods
using different age-tracers such as Cepheids, planetary nebulae, and open clusters in the Milky Way \citep{Andrievsky02,Maciel03,Maciel10}. 
Direct observations of metallicity gradients at high redshift have not been possible until the recent employment of  high sensitivity  
near infrared (NIR) integral field spectrographs (IFS) on large telescopes.  
  
As a result, it is only in the past three years that the first radial metallicity gradient measurements have been made for high-redshift galaxies \citep{Jones10b,Cresci10,Yuan11,Queyrel12,Swinbank12a,Jones12}. These integral field unit (IFU) studies can be divided into three categories according to their angular resolution:  (1) gravitationally lensed galaxies with adaptive optics (AO) aided observations \citep{Jones10b,Yuan11,Jones12}; (2)  non-lensed galaxies with AO aided observations \citep{Swinbank12a}; (3) non-lensed galaxies with seeing-limited observations \citep{Cresci10,Queyrel12}.  It is interesting to note
that  the steepest gradients discovered so far  are all observed in the highest resolution category  (1), whereas much shallower gradients are reported in  categories (2) and (3).

Compared to local observations,  the most obvious restrictions at high redshift are  angular resolution and signal-to-noise (S/N). 
Because of these impediments, simplifications in data analysis such as annular binning are used to gain S/N in spectra.  It is unknown whether/how
 the observational limitations and data analysis techniques 
for high redshift studies  cause systematic  uncertainties in the measured metallicity gradients. 

With the rapid progress in metallicity gradient observations at high redshift,  it is imperative that we understand the systematics in our measurements
before embarking on surveys of large samples.

This paper is the first of a series devoted to understanding the systematics of  metallicity gradient measurements.
Specifically,  this paper investigates how the metallicity gradient is affected by the angular resolution, S/N, and annular binning that are characteristic for high-$z$ studies.

We use two approaches:  1. We downgrade the highest angular resolution and S/N IFU data 
from our published AO-aided lensed galaxy to lower resolutions and then 
recalculate  the metallicity gradients at different resolutions; 2. We simulate the 
IFU data at high-$z$ using  a local starburst galaxy with a known gradient. We then compare the metallicity gradients measured from the simulated data at different  angular resolutions 
and S/N  ratios. 
We find that the metallicity gradient changes systematically with angular resolution, S/N and the annular binning parameters. 

This paper is organized as follows.
In Section~\ref{assumption}, we  list the assumptions used.
In Section~\ref{sec:secA} and Section~\ref{sec:secB}, we describe our two approaches and results. 
We compare our results with literature data in Section~\ref{sec:lit}.
We  discuss  our explanations for the systematic uncertainties in Section~\ref{sec:dis}.
We present our conclusions and future directions  in Section~\ref{sec:sum}.
Throughout this paper we use a standard $\Lambda$CDM cosmology with $H_0$= 70 km s$^{-1}$
Mpc$^{-1}$, $\Omega_M$=0.30, and $\Omega_\Lambda$=0.70.

\section{Overall Assumptions}\label{assumption}
Many factors can affect metallicity gradient measurements. 
In order to analyze the role of  angular resolution, S/N and annular binning parameter, 
it is necessary to decouple these effects from other factors that may affect the metallicity gradient.  
We hold all other parameters that may affect the gradient constant.  
 To focus on the systematics of the angular resolution, S/N and annular binning on the metallicity gradient measurement, 
we find it necessary to impose the following assumptions and simplifications:

\begin{itemize}
\item  We use the \nii\,/\ha\ ratio as calibrated in \citet{Pettini04} (the PP04N2 method hereafter) to calculate metallicity.  Because of the relatively easy access,  \nii\,/\ha\ ratios will continue to be the most practical metallicity diagnostics for high-$z$ studies in the near future.
In \citet{Kewley08}, we showed that the absolute abundance scale differs for different calibrations.  However, relative metallicities, such as metallicity gradients are robust to within 0.03 dex on average.  To avoid problems with the absolute abundance calibration scale, we only apply the single PP04N2 diagnostic.  We have verified that the results for our local galaxy are unchanged when independent diagnostics such as the \citet{Kobulnicky04} and \citet{McGaugh91} calibrations are applied.

\item We assume that the metallicity gradient can be fitted by  one linear function \citep[c.f.,][]{Bresolin09}. We express gradients in dex per kpc.  We 
do not consider how the intrinsic sizes of galaxies  may affect the metallicity gradient measurement \citep[e.g.,][]{Jones12}. 

\item We assume that the metallicity distributions in disk galaxies have cylindrical symmetry, i.e.,  azimuthal abundance variations are not significant \citep[e.g.,][]{Kennicutt96, Bresolin09b}. 
 
\item  We assume the galactic center is well-defined and is aligned with the peak of \ha\ emission. Locating the galactic center is straightforward for local galaxies, but may not be unambiguous
for high-$z$ galaxies because of their irregular morphology \citep[e.g.,][]{Abraham96,Giavalisco96}.   For similar reasons, we do not correct for the inclination angle of the galaxy as it is not well constrained for high-$z$ galaxies.   Our aim is to simulate high-redshift data and to analyze that data under the same assumptions and limitations that affect high redshift galaxy observations.

\item  We only consider metallicity gradients for isolated disks.

\item The inhomogeneity of metal distribution has been found in both the radial and vertical directions of the galactic discs, as well as in early-type galaxies. 
Although abundance gradients in the vertical direction and  in early-type galaxies are equally important \citep{Franx90,HenryR99,Marsakov06}, 
we confine our discussion to  the radial gas-phase metallicity in disk galaxies.
 
 \item The  uncertainties stemming from lensing models and the effects of shocks and AGN \citep[e.g.,][]{Wright10a,Rich11,Yuan12a,Westmoquette12}
  are beyond the scope of the paper and will be investigated in a future paper in this series. 

\end{itemize}

Note that each of the above items may play an important role in metallicity gradient studies. We hold them constant  in order to   
filter out the systematics of the angular resolution, S/N and annular binning on the metallicity gradient measurement. 
By comparing metallicity gradients obtained for the same data within the same galaxy using the same calibration, 
we avoid the effects of the issues highlighted above.

\section{Methodology A: Downgrade High-Redshift High Angular Resolution Data}\label{sec:secA}
The highest intrinsic angular resolution  achieved ($\sim 0.02^{\prime\prime}$) in metallicity gradient measurements at high-$z$  is in category (1):  gravitationally lensed galaxies with adaptive optics (AO) \citep{Jones10b,Yuan11,Jones12}.    Among these studies,  the grand-design spiral Sp1149 at $z=1.49$ is most spectacular and suitable for 
detailed metallicity analysis because of its clear face-on morphology and fortuitously uniform lensing magnification \citep{Yuan11}.  
Note that all high-$z$ observations of  galaxies suffer from the loss of low surface brightness pixels.
 The magnification of gravitationally lensed galaxies helps to alleviate the problem by
bringing more spatial elements with low surface brightness within the detection limit compared to non-lensed cases.
We therefore use Sp1149 as our high redshift testing galaxy.

\subsection{Integral field spectroscopy of Sp1149}
\begin{figure}[!ht]
\begin{center}
\includegraphics[trim = 3mm 3mm 3mm 5mm, clip, width=8.8cm,angle=0]{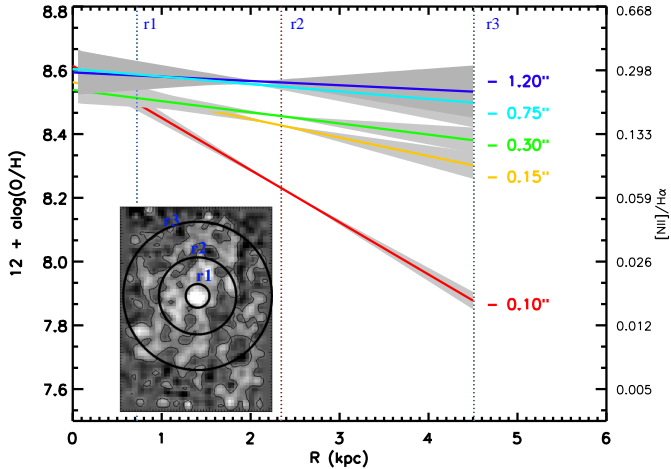}
\caption{Metallicity (\nii\,/\ha\ ratio) vs. radius for  Sp1149 downgraded to different angular resolution FWHM = 0.1, 0.15,  0.3,  0.75,  and 1.2 arcsec (in different colors). 
The solid lines are the best-fit of the metallicity slope, the 1-$\sigma$ errors of the slope from linear regression are shown as shaded regions around the solid lines. 
The imbedded panel on the lower left shows the original OSIRIS \ha\ map with the three annuli marked.   The original data
has an angular resolution of  0.1$^{\prime\prime}$ (FWHM), corresponding to the pixel scale of the  data.  
Note that because of lensing magnification, the corresponding intrinsic angular resolution
in this case is 0.02$^{\prime\prime}$.
}
\label{fig:sp1149FWHM}
\end{center}
\end{figure}

Sp1149 was observed  using the Laser Guide Star Adaptive Optics aided integral field spectrograph  (OSIRIS; \citealt{Larkin06}) on KECK II in 2011.  We 
 achieved an image-plane angular resolution of  $0\farcs1$, corresponding to a  source-plane resolution of  $\sim$ 0.02$^{\prime\prime}$ (170 pc). The pixel scale 
we use for Sp1149 is $0\farcs1$, with a field of view (FOV) of 4.8$^{\prime\prime}$  by 6.4$^{\prime\prime}$ and a spectral resolution of $R\sim$3400.
 
Sp1149 is a rare lensing case as it is stretched almost equally by $\sim$ 5 times on each side  of the two-dimensional image. As a result, the source-plane morphology looks
almost identical to the image-plane morphology.  The error on the lensing magnification of Sp1149 is $< 10\%$ \citep{Smith09,Yuan11}. 

In   \citet{Yuan11}, we  derive the metallicity gradient for Sp1149 by integrating
 the spectrum in three annuli corresponding to a physical length of    r1$=$0.72$\pm$0.1 kpc, r2$=$2.34$\pm$0.2 kpc, r3$=$4.5$\pm$0.4 kpc.  The choice of the annuli is such that the integrated spectra could reach S/N $>$ 5 for the weak \nii\ lines.  The \nii\ line is robustly detected at S/N $>$ 5 for the inner two annuli and is a 3$\sigma$ detection for the outer annulus which we give an upper limit.  In the next subsection, we use the same steps to derive metallicity gradients on the resolution-downgraded data of Sp1149.

\subsection{Sp1149 Downgraded to Different Angular Resolutions}
For each wavelength slice, we convolve the spatial pixels with a Gaussian kernel  of  a range of full width half maximum (FWHM) values (0.2$^{\prime\prime}$-0.8$^{\prime\prime}$).
We then re-extract the spectra using the same three annuli  as  in \citet{Yuan11}.  We use the same line-fitting  procedures as used in \citet{Yuan11}: 
Gaussian profiles were fitted simultaneously to the three emission lines:  \nii\lam6548, 6583 and \ha. 
The line profile fitting was conducted using a $\chi^2$ minimization procedure which takes into account the greater noise level close to atmospheric OH emissions.
The centroid and velocity width of \nii\llam6548, 6583 lines were constrained by the velocity width of \ha\lam6563, and 
the ratio of \nii\lam6548 and \nii\lam6583 is constrained  to  the theoretical value  \citep{Osterbrock89}. 
We fit the \nii\,/\ha\ ratios in the  three annuli and obtain the metallicities using the PP04N2 method \citep{Pettini04}.  
In Figure~\ref{fig:sp1149FWHM} we show the metallicity gradient and 1-$\sigma$ error of the fit to metallicity (\nii\,/\ha\ ratio) vs. radius for the downgraded IFU data
in a few cases of angular resolution FWHM.
 
We see from Figure~\ref{fig:sp1149FWHM} that  the measured metallicity gradient flattens with poorer angular resolution FWHM. 
With  an angular resolution FWHM  above $\sim$0.75$^{\prime\prime}$, inverted (positive) gradients begin to appear within the  errors of the linear fitting. 

\begin{figure}[!ht]
\begin{center}
\includegraphics[trim = 5mm 5mm 5mm 5mm, clip, width=6cm,angle=90]{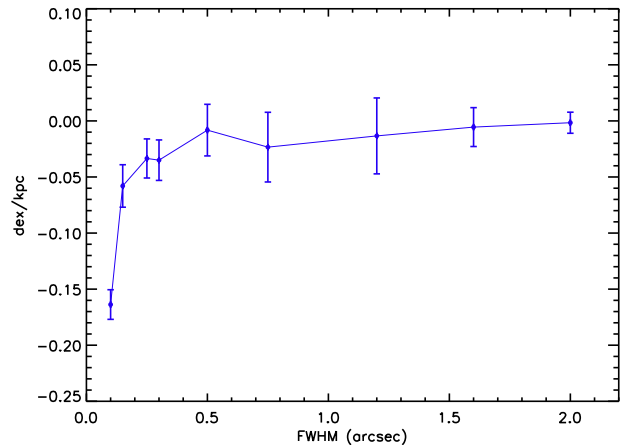}
\caption{The measured metallicity gradient and the 1-$\sigma$ error as a function of  the FWHM angular resolution based on the data of Sp1149 at $z=1.49$ \citep{Yuan11}. 
The smallest angular resolution corresponds to the actual observed resolution of 0.1$^{\prime\prime}$ or 0.02$^{\prime\prime}$ intrinsic resolution after correcting for lensing magnification.
}
\label{fig:curvesp1149}
\end{center}
\end{figure}

To determine whether the  trend continues at even lower angular resolution,   we downgrade  the data using a range of angular resolution FWHM between the original 0.1$^{\prime\prime}$ resolution and a fiducially  large 2.0$^{\prime\prime}$ resolution.  We find that relation between  the  metallicity gradients and the angular resolution FWHM follows an interesting curve (Figure~\ref{fig:curvesp1149}).

\begin{figure}[!ht]
\begin{center}
\includegraphics[trim = 2mm 0mm 1mm 0mm, clip, width=6.cm,angle=0]{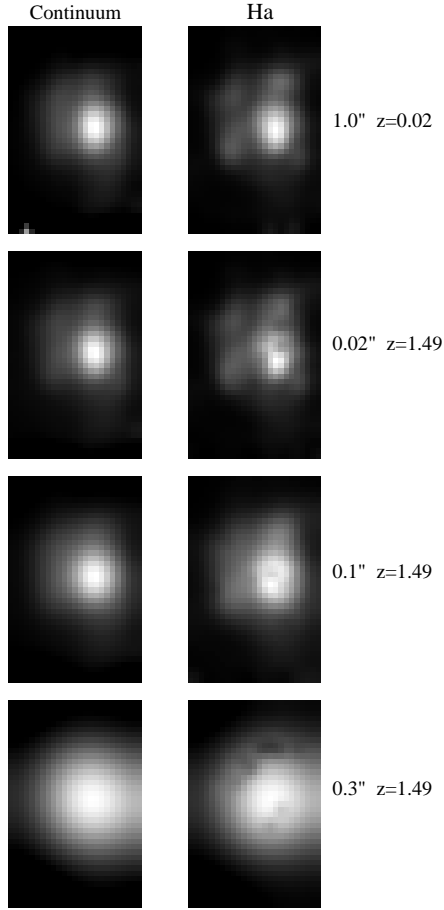}
\caption{The continuum and wavelength-collapsed  \ha\  image derived from the WiFeS data of Sp17222 (first row), and from the simulated Sp17222 at $z=1.49$
in different angular resolution FWHM (row 2-3: examples of FWHM \,=\, 0.1$^{\prime\prime}$, 0.2$^{\prime\prime}$, 0.3$^{\prime\prime}$).  We see that 
at FWHM $\ge$ 0.3$^{\prime\prime}$, the \ha\ clumps on the outer disk of Sp17222 are not resolved anymore. The artifacts (black dots) on the simulated data are regions
of low S/N ($<$ 5) on the \ha\ line.
}
\label{fig:simuHa}
\end{center}
\end{figure}
\begin{figure*}[!ht]
\begin{center}
\includegraphics[trim = 30mm 0mm 4mm 6mm, clip, width=16cm,angle=0]{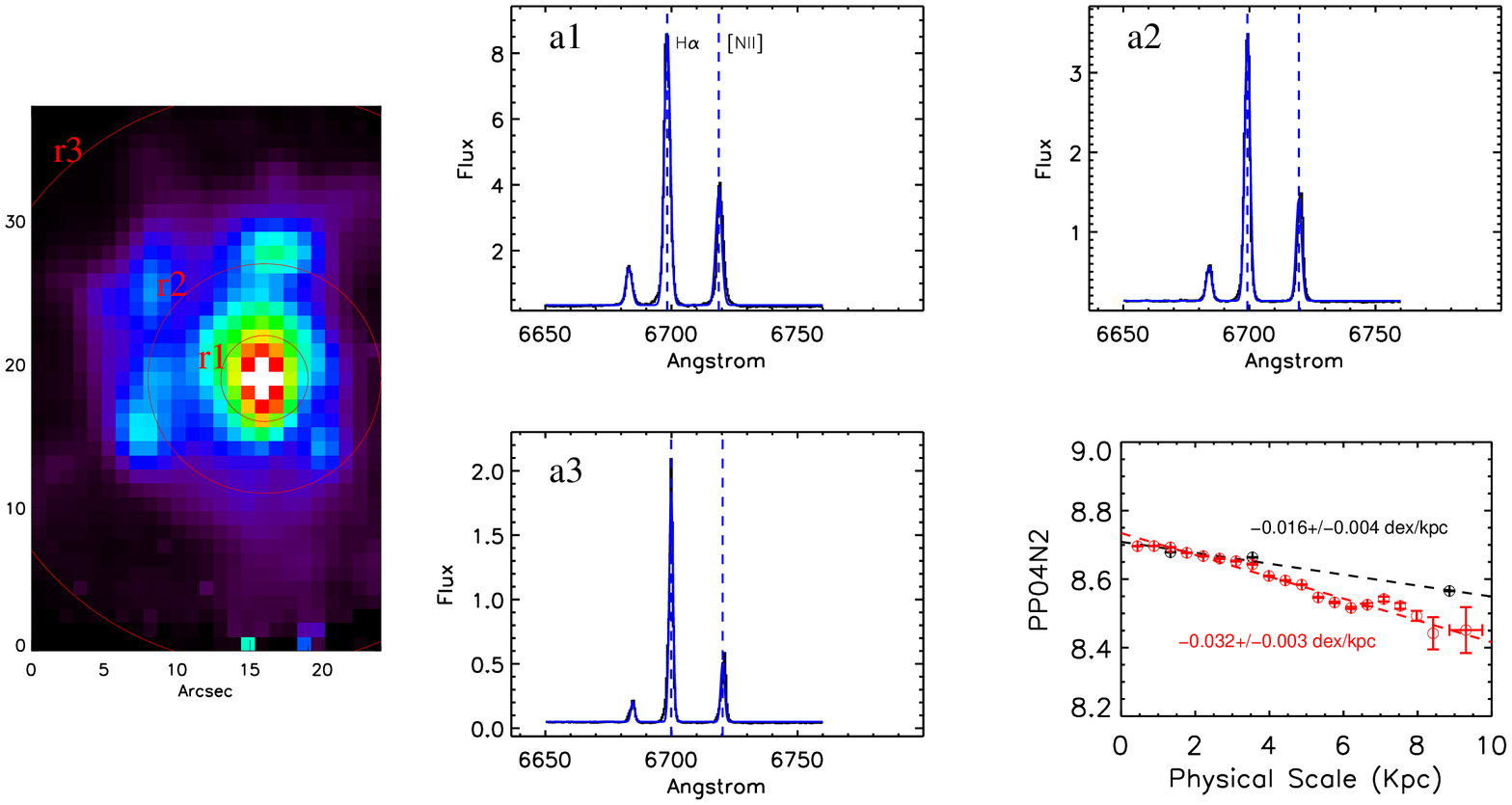}
\caption{Derived \nii\,/\ha\,-based metallicity gradient for the local data of Sp17222.  
Left: \ha\ intensity derived from  the WiFeS datacube. We use the same three annuli as defined on the simulated high-$z$ data. 
The average spectra (zoomed in for the \nii\,and \ha\, vicinity) in the three annuli are shown on the right panel (a1-a3). 
The flux is in unit of $10^{-16}$~ergs~s$^{-1}$~cm$^{-2}\AA^{-1}$.  
The panel on the bottom right shows the metallicity gradient derived using the three annuli (black) in comparison with a full sampling (red) without binning.
}
\label{fig:local1}
\end{center}
\end{figure*}

Figure~\ref{fig:curvesp1149} suggests that there is a critical angular resolution, below which the measured gradient is significantly more steepend.  Above this critical resolution, e.g., under  seeing-limited conditions, the gradient  approaches a very shallow or  nearly flat slope.   

The behavior of the gradient vs. angular resolution is  worrisome.  Ideally, the measured gradient should not be a function of angular resolution within the observational errors.
If a higher angular resolution more closely represents  the real gradient, then what is the angular resolution required to recover the intrinsic gradient ?  We address this question in 
Section~\ref{sec:secB}.

\section{Methodology B: Simulate High-Redshift IFU Observation Using Local Data}\label{sec:secB}
\subsection{Local Spiral Galaxy IRAS F17222-5953 (Sp17222)}
In order to simulate an IFS observation at high-$z$, we choose a local galaxy that has star formation concentrated at the galactic nucleus instead of in the spiral 
arms.   We use the isolated luminous infrared galaxy IRAS F17222$-$5953 (Sp17222) at $z=0.021$. Sp17222 is relatively face-on, has a morphological type of Sbc, 
and a moderate SFR (H$\alpha$) of $\sim$11 M$_{\odot}$~yr$^{-1}$.  The optical   IFU data of Sp17222 is adopted from the Wide Field Spectrograph (WiFeS) Integral Field Unit (IFU) Great Observatory All-Sky LIRG Survey (GOALS) Sample (WIGS) \citep{Rich12}.

Briefly, the IFU data were taken using WiFeS at the Mount Stromlo and Siding Spring Observatory (SSO) 2.3 m telescope \citep{Dopita10}.
The blue and red spectra of WiFeS were observed  with a resolution of R3000 and R7000 and wavelength coverage of $\sim$3500-5800 $\AA$  and $\sim$ 5500-7000 $\AA$ respectively.
We use the fully reduced and flux calibrated data from \cite{Rich12}.  
The pixel scale of WiFeS is 0.5$^{\prime\prime}$, with a field of view (FOV) of 25$^{\prime\prime}$$\times$38$^{\prime\prime}$.  The reduced data are 
binned by 2 pixels in the spatial direction, yielding an effective spatial resolution of 1.0$^{\prime\prime}$$\times$1.0$^{\prime\prime}$.

\subsection{Simulating IFU Observations of Sp17222  at  $z=1.49$}\label{sec:simuobs}
To make realistic  comparisons with observed IFU data, we fix Sp17222 at the redshift of  Sp1149 ($z=1.49$). We derive the size and surface brightness distribution for each redshifted 
wavelength slice using basic equations of angular diameter distance and luminosity distance, and based on the conservation of total luminosity. 
Due to pure cosmology effects, the angular size and total flux of Sp17222  are reduced by $\sim 0.05$ and $\sim 2.9\times10^{-5}$ respectively at 
$z=1.49$.  

Because the angular size becomes smaller at higher redshift, an angular scale of 0.02$^{\prime\prime}$ is required to fully recover the spatial samplings of the local IFU data. 
We thus set the highest FWHM resolution of our simulation as 0.02$^{\prime\prime}$.  

The original Poisson-noise dominated optical spectra are shifted into the NIR at $z=1.49$, 
and real NIR data are sky-background dominated. We  compose a noise spectrum at the spectral resolution of WiFes 
by interpolating  the sky-residuals of our OSIRIS datacube. The noise spectrum represents the sky OH emission residuals in a real observation in the NIR at $z=1.49$. 
The noise spectrum is scaled and added to the redshifted spectrum of each spaxel according to the required input S/N. 

We define the input S/N of our simulation on the spaxel where \ha\ emission peaks.  For example, an input S/N of 100 means that the noise spectrum is 
scaled and added to the target spectrum such that the S/N ratio of the \ha\ line on the brightest pixel is 100.
To facilitate comparison with observations,  
we also  measure the observed S/N of the \nii\ lines in outer annulus on the simulated datacube (Section 4.5). 

Note that in this simulation, we do not consider any intrinsic evolutionary effect except that we multiply the total flux of the original IFU data by a factor of 20.
The factor of 20 is adopted for two reasons:  1. The SFR of galaxies at $z \sim 2 $ are found to be $\sim 20$ higher than galaxies at $z \sim 0$ at a fixed mass
of $\sim 10^{10}~M{_\odot}$ \citep[e.g.,][]{Noeske07b,Noeske07a,Zahid12};  2. Without manually increasing the total flux, Sp17222 would have been under the detection limit of any current NIR IFU instrument.  We find that by increasing the intrinsic flux by 20, Sp17222 would be well detected in NIR IFU spectrographs such as OSIRIS on KECK. The factor of $\sim$ 20 also
coincides with the lensing flux magnification of Sp1149 in Section~\ref{sec:secA}.

Finally, following similar steps as in Section~\ref{sec:secA}, we convolve the  spatial pixels with different angular resolution FWHM and generate a set of simulated IFU data at $z=1.49$. 

For both the local and simulated datacube, we use the same line-fitting procedures as described in \citet{Yuan11} and in Section 3.2.
Figure~\ref{fig:simuHa} shows  the continuum and \ha\ emission line image derived from the original WiFeS data of Sp17222, and from the simulated high-$z$ Sp17222 datacube 
in different angular resolutions.

\subsection{The Effect of Annular Binning on Metallicity Gradient}

\begin{figure}[!ht]
\begin{center}
\includegraphics[trim = 0mm 0mm 0mm 0mm, clip, width=6.5cm,angle=90]{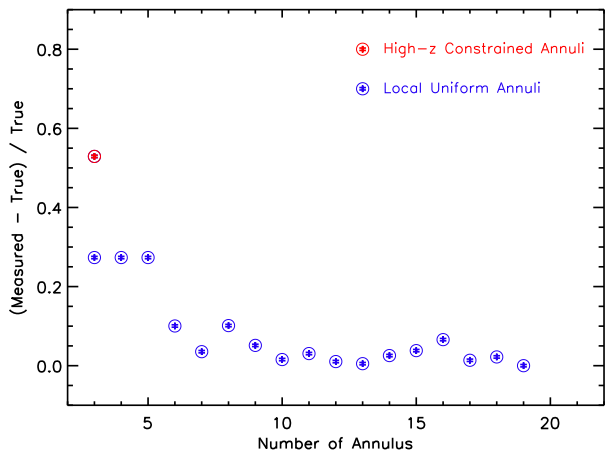}
\caption{The percentage of the difference between the measured metallicity gradient in annuli and the true metallicity gradient in full sampling as a function
of the number of annulus used. The blue circles are the measurements from the annuli that are uniformly divided from the galactic center to the outer disk. 
The red circle shows the measurement from the 3 annuli that are defined on the high-z data constrained by the S/N of the weak \nii\ lines (Section 4.3).
}
\label{fig:rec}
\end{center}
\end{figure}

Annular binning is a commonly used method to derive metallicity gradients for low S/N high-$z$ data \citep{Jones10b,Yuan11,Queyrel12,Swinbank12a}.
In order to quantify the effect of annular binning on metallicity gradient measurements, we compare the effect of 3-annuli binning  with full sampling without binning.
Since only the local datacube has sufficient spatial resolution and S/N simultaneously  to compare a full and 3-annuli sampling, 
we test these 2 types of samplings on the original datacube of Sp17222 first, without degrading the resolution.

We show the metallicity gradients derived from a 3-annuli binning and a full sampling in radius in the lower right panel of 
Figure~\ref{fig:local1}.  The three annuli are defined on the simulated high-$z$ data and they are chosen
 such that the \nii\ line from the outer-most annulus is robustly detected (S/N $\ge$ 5) on all annuli of the  high-$z$ datacube, 
 consistent with current high-z studies  \citep{Jones10b,Yuan11}.

We see from Figure~\ref{fig:local1} that the gradient from annular binning is shallower than the gradient derived from full samplings. 
In the case of Sp17222,  binning in 3 annuli yields a metallicity gradient of -0.016$\pm$0.004 dex kpc$^{-1}$, whereas the fulling sampling
yields -0.032$\pm$0.003 dex kpc$^{-1}$.  

To further investigate the relation between the measured gradient and the number of annuli used, we calculate the measured metallicity
on the local IFU data of Sp17222 using a range of annuli.  Note that the  annuli in Figure~\ref{fig:local1}
are defined on the simulated high-$z$ data and are constrained by  the S/N of  the \nii\ line from the outer-most annulus. 
The choice of the annuli is therefore defined on the emission line flux distribution of the high-$z$ data. 
A more straight-forward  method of defining annuli is to divide the radius equally into N uniform bins. 
These uniform annuli can be applied to the local data where  the requirement of S/N$>5$ on  the \nii\ line is not a constraint for most radii. 
We therefore divide the Sp17222 data equally in radius using a number of  N$_{min}$\,=\,3 and N$_{max}$\,=\,20 annuli. 
We show in Figure~\ref{fig:rec} the percentage of the difference between the measured metallicity gradient from different numbers of annuli and the true metallicity gradient from full sampling
as a function of the number of annuli.
We find that with N=3-5 uniform annuli, the measured gradients are close to the true gradient within $\sim$ 30\%; with N$\ge$6 uniform annuli, 
the true gradient can be recovered to within 10\%.  We also show that the metallicity gradient from the uniform 3-annuli is closer to the true gradient (28\% c.f. 54\%) than
the S/N constrained 3-annuli employed on the high-z data. 

In summary,  we see that binning in 3 (or another small number of) annuli introduces non-negligible systematic errors on the metallicity gradient measurement. 
The definition of the location of the annuli also plays a non-negligible role in the metallicity gradient measurement.  The choice of the annuli used in high-z 
studies is weighted by the S/N of the weak \nii\ lines and can only recover the true gradient to within 54\% in the simulated case of Sp17222. 
The choice of uniformly distributed annuli recovers the true gradient significantly better. In the case of Sp17222, using N $>$ 6 annuli can recover 
the true gradient to within 10\%.

\begin{figure*}[!ht]
\begin{center}
\includegraphics[trim = 0mm 0mm 0mm 8mm, clip, width=10cm,angle=90]{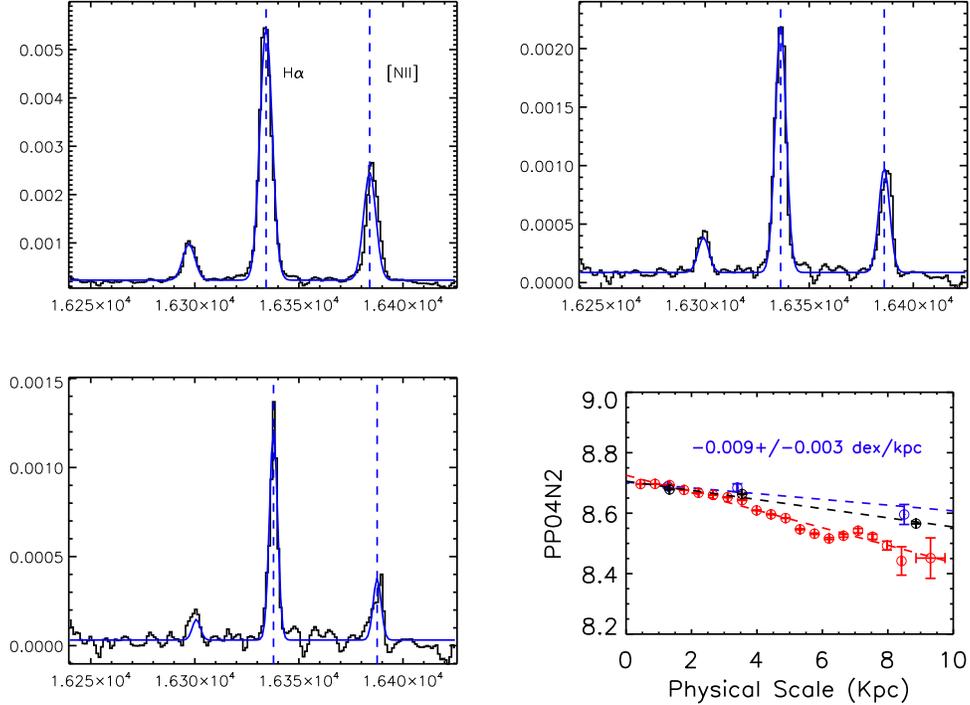}
\caption{Examples of the \nii\ and \ha\ lines extracted from the 3 annuli on a simulated datacube (S/N=100, FWHM=0.02$^{\prime\prime}$).  The panels are  
organized in the same order as in Figure~\ref{fig:local1}, with the same flux and wavelength units.  
The panel on the bottom right shows the metallicity gradient (blue) based on the  \nii\ and \ha\  line ratios. The
local 3-annuli (black) and full sampling (red) gradients are also shown for comparison.
}
\label{fig:fitspec}
\end{center}
\end{figure*}

\begin{figure*}[!ht]
\begin{center}
\includegraphics[trim = 10mm 4mm 4mm 8mm, clip, width=18cm,angle=0]{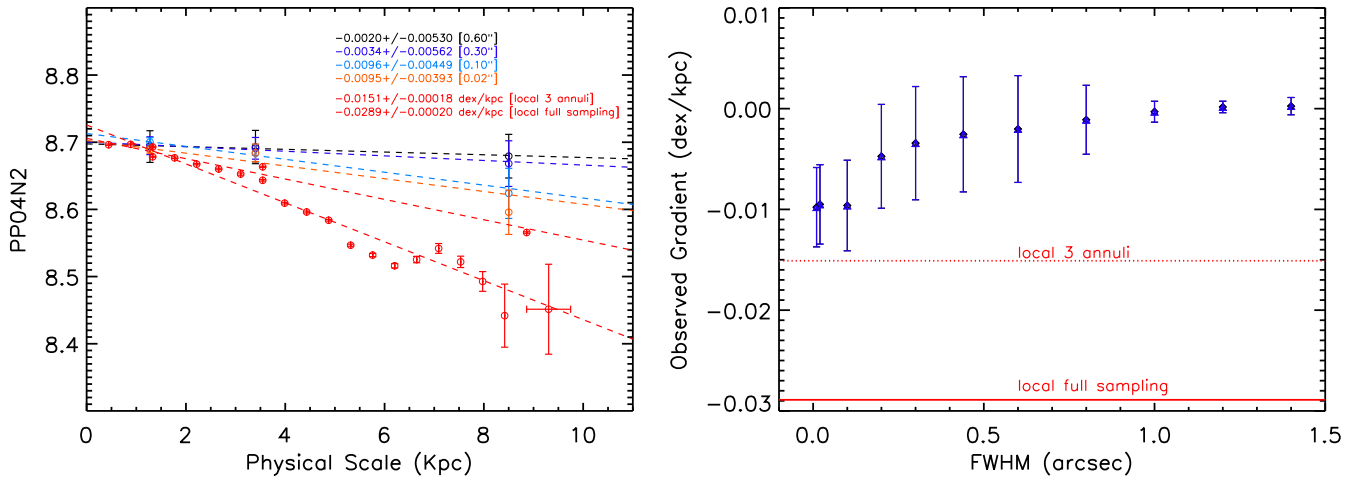}
\caption{Left:  Metallicity vs. radius on the simulated datacube (S/N fixed at 100).  Three-annuli  bins are used. We show examples of a few angular resolution FWHM in different colors.
For comparison, the three annuli and fully sampled metallicity of the local data are shown in red. Right: Simulated metallicity gradient as a function of  angular resolution FWHM.
The horizontal red lines show the location of the metallicity gradients measured on the local data using three annuli and full samplings respectively.
We see that the local gradient is not recovered even in the highest
resolution of the simulation (FWHM=0.02 $^{\prime\prime}$). 
}
\label{fig:simugrad}
\end{center}
\end{figure*}

\begin{figure}[!ht]
\begin{center}
\includegraphics[trim = 8mm 8mm 5mm 8mm, clip, width=6.25cm,angle=90]{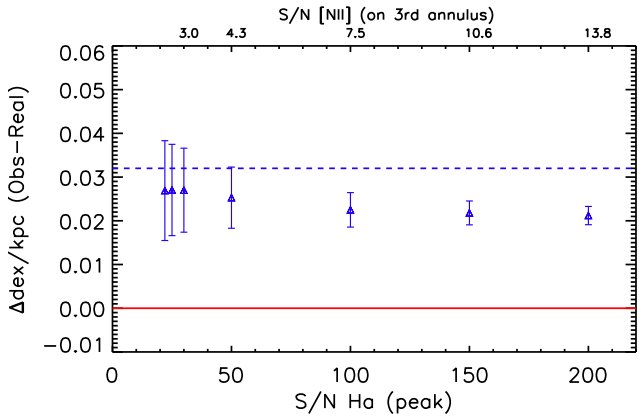}
\caption{The deviation of the measured gradient from the ``true gradient" as a function of S/N (angular resolution fixed at 0.02$^{\prime\prime}$). 
The horizontal red line shows the true gradient measured on the 
fully sampled local data of Sp17222.  The S/N marked on the bottom are defined on the peak spaxel of the \ha\ emission line (Section 4.2).  The 
corresponding  S/N of the \nii\ line on the faintest annulus are marked on the top.
}
\label{fig:gradSN}
\end{center}
\end{figure}

\begin{figure}[!ht]
\begin{center}
\includegraphics[trim = 8mm 8mm 5mm 8mm, clip, width=6.35cm,angle=90]{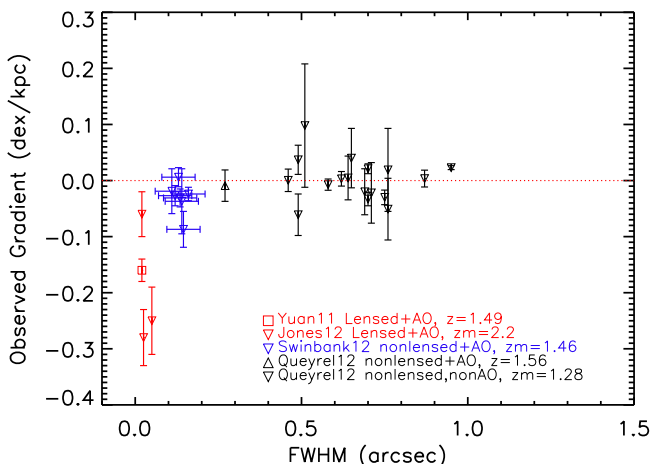}
\caption{The fact that  measurements for different galaxies in different studies follow the same trend as Figure 2 and 7 strongly suggest 
that the angular resolution has played a role in causing the systematic difference in the metallicity gradient. The effect of angular resolution
needs to be quantified before interpreting the metallicity gradient evolution with redshift. 
}
\label{fig:lit}
\end{center}
\end{figure}

\subsection{The Effect of Angular Resolution on Metallicity Gradient Measurements}\label{sec:fwhmsimu}
We extract the 3-annuli spectra for our simulated high-$z$  datacube with a range of  angular resolution FWHM and S/N ratios. 
Figure~\ref{fig:fitspec} shows an example of the 3-annuli spectra and derived metallicity gradient.  The example in Figure~\ref{fig:fitspec}
has an angular resolution of FWHM=0.02$^{\prime\prime}$ and a limiting S/N of 100 on the  \ha\ line.
A S/N of 100 on the \ha\ line corresponds to a measured S/N of 7.5 on \nii\ line of the faintest annulus. 

Figure~\ref{fig:simugrad} shows the PP04N2-based metallicity gradient for different angular resolution FWHM.
We see the same trend in the simulated data as seen in Figure~\ref{fig:sp1149FWHM}: the measured metallicity gradient is shallower as the FWHM increases.
As the angular resolution FWHM approaches the seeing limited regime, the gradients are essentially consistent with a flat ($\sim$ 0) value.
The curve of the metallicity gradients vs. the angular resolution  FWHM relation is similar to Figure~\ref{fig:curvesp1149}.  However, 
the  change of the FWHM from steep to shallow gradients is slower.

The deviation of the measured gradient from the true gradient measured 
from the fully sampled local data of Sp17222 can be derived from Figure~\ref{fig:simugrad}. 
The closest  gradient measurement we can obtain is $0.095\pm0.004$ dex per kpc (true gradient =  0.032$\pm$0.004 dex per kpc), i.e., 
only 29\% of the true gradient is recovered by the highest angular resolution and S/N observation simulated.

\subsection{The Effect of S/N on Metallicity Gradient Measurements}\label{sec:secsn}
Section~\ref{sec:fwhmsimu} shows the high input S/N (100) case.  As described in Section~\ref{sec:simuobs}, 
the input S/N is defined on the \ha\ line of the central pixel. 
 
We repeat our analysis in  Section~\ref{sec:fwhmsimu}  using a range of input S/N. We vary the S/N ratios as described in Section~\ref{sec:simuobs}. 
We fix the angular resolution at FWHM=0.02$^{\prime\prime}$.
Figure~\ref{fig:gradSN} shows the deviation of the measured gradient from the  true gradient as function of S/N. 
The errors of the metallicity gradient are derived from the errors in the slope of the two-variable (metallicity and radius) linear regression.
 We see that the most significant effect of  S/N is on the errors of the metallicity gradient measurement. 
 The errors of the gradients decrease with S/N. At  \ha\ S/N (peak) $<$ 50 or \nii\ S/N  (3$^{rd}$ annulus) $< $5, the error bars of the gradient are so large that 
 the gradients can  not be meaningfully constrained, i.e., both a positive and negative gradient are consistent within the errors.

\section{Metallicity Gradient vs. Angular Resolution from the  Literature}\label{sec:lit}
As described in Section 1, current IFS studies on metallicity gradients can be divided into three categories according to their angular resolution:  (1) gravitationally lensed galaxies with AO; (2)  non-lensed galaxies with AO; (3) non-lensed galaxies under seeing-limited conditions. 
 We gather the literature data in these 3 categories and examine the measured metallicity gradients with respect to the angular resolution FWHM used in the observations.

For category (1), we use the 5 lensed galaxies from \citet{Jones10b,Yuan11,Jones12}. We exclude 1 merging system from the lensed galaxies
because galaxy mergers are known to flatten metallicity gradients \citep{Kewley10,Rupke10b,Rich12}.  
For (2), we use the 7 AO-corrected galaxies
 from \citet{Swinbank12a}.  The galaxy sample in  \citet{Swinbank12a} is chosen from the
 High-Z Emission Line Survey (HiZELS) survey \citep{Sobral12b}  which targets \ha\ emitters  close to bright stars.
   For (3), we use the data of \citet{Queyrel12}, as 17 out of the 18 galaxies in \cite{Queyrel12} are seeing-limited measurements, with 1 AO corrected measurement.  
 The galaxy sample in  \citet{Queyrel12} is part of the MASSIV project \citep{Contini12} and 
 is chosen based on the visibility of the \ha\ line in J or H band. 
These 18 galaxies do not include interacting galaxies identified in  \citet{Queyrel12}.
Since  \nii\ and \ha\  lines are available for all the samples, we calculate the metallicities using the PP04N2 metallicity diagnostic. 
Note that the metallicity gradients  in  \citet{Queyrel12} are based on the  \citet{Perez09b} \nii\,/\ha\ metallicity calibration,  which we have converted to the PP04N2 metallicity calibration. 
The metallicity gradients are shown as a function of the angular resolution FWHM in Figure~\ref{fig:lit}. 

Note that there is a redshift difference among the three samples.  The median redshift for sample (1) is $z\sim2$, the 
median redshifts for samples (2) and (3) are $z\sim1.5$ and $z\sim1.3$ respectively.  If the angular resolution effect does not play a role in these observations, 
one would be tempted  conclude that the metallicity gradient is evolving from the steeper gradients at $z\sim2$ to much shallower gradients at $z\sim 1$. 
However, relating Figure~\ref{fig:lit} with Figure~\ref{fig:curvesp1149} and Figure~\ref{fig:simugrad},  we see a striking similarity in the behavior of the measured metallicity gradient
vs. the angular resolution FWHM relation.   The fact that the steepest gradients are only seen in the lensing+AO samples (including one galaxy at z$\sim$ 1.5) is consistent with our findings 
in the previous sections that there exists a critical angular resolution limit beyond  which the absolute value of the gradients are  significantly under-estimated.
It is possible that angular resolution  causes the trend  in Figure~\ref{fig:lit}. 
Until the effect of angular resolution on metallicity gradients is well understood, extreme caution must be taken when interpreting apparent evolution in metallicity gradients with redshift.

\section{Discussion}\label{sec:dis}
\subsection{Explanation of the angular resolution effect: ``low S/N line smearing" effect}\label{sec:exp}
The PP04N2 based metallicity depends on the ratio of the \nii\, and \ha\ lines. The \nii\ lines are usually more than 3 times weaker than \ha\ lines for normal star-forming regions.
Given a negative radial metallicity gradient, much weaker \nii\, lines are expected in the outer regions of a galaxy.  Any spatial averaging/smoothing process
selectively smears the low S/N \nii\, line regions. As a result, the spatially averaged
spectra are weighted more towards the regions of stronger \nii\, lines, leading to over-estimation of the metallicity in the outer-disk. 
Accordingly, binning using small numbers  of annuli has the same effect as lower resolution (i.e. larger angular resolution FWHM). 

Using this argument, we would expect that the steeper the intrinsic gradient is, the larger the observed gradient is going to deviate from the intrinsic value. 
The curve of the gradient vs. FWHM relation (e.g., Figure~\ref{fig:curvesp1149}, Figure~\ref{fig:simugrad}, and  Figure~\ref{fig:lit}) would depend on the intrinsic shape of the metallicity
gradient.  

If low S/N line smearing is the cause of the angular resolution effect observed in  Figure~\ref{fig:curvesp1149}, Figure~\ref{fig:simugrad}, and  Figure~\ref{fig:lit}, one would think that it is possible to cancel out the smearing effect by using a reversely weighted function. 
However, it is not straightforward to establish this anti-weighting function, as it relies on knowing the intrinsic gradient  a priori.  Moreover, if the ``smearing" process is done by the atmosphere, it is impossible 
to de-convolve the  observed FWHM back to the pre-smeared values, as it would be equivalent to finding an algorithm to  ``recover" the data with any wished angular resolution.

\section{Conclusion}\label{sec:sum}
In this paper, we have demonstrated that the angular resolution, S/N and annular binning introduce significant
systematic errors in metallicity gradient measurements at high-$z$. 

We find that: 
\begin{itemize}

\item  The measured metallicity gradient changes systematically with angular resolution FWHM.  Seeing-limited observations are likely to produce
more flattened gradients than AO-aided high-resolution studies.

\item  There is a critical angular resolution FWHM range beyond which the measured metallicity is  significantly more flattened than the intrinsic metallicity. This critical FWHM 
depends on the intrinsic gradient of the galaxy.  For the two cases used in this work, the critical FWHM is $< 0.02^{\prime\prime}$, only currently reachable with AO + gravitational lensing.

\item  For a fixed angular resolution, the errors of the metallicity gradient increase as S/N decrease. At low S/N ($< 5$ for \nii\ line in the simulated case), the errors 
are so large that the gradient can not be meaningfully constrained.  

\item  Three-annuli binning or any limited number of annular binning yields a more flattened gradient than the intrinsic gradient.  

\end{itemize}

Until these effects are  thoroughly understood, we urge caution in interpreting metallicity gradient evolution with redshift.
Our next work is to build an empirical angular resolution  library by simulating a large sample
of high-z galaxies using local galaxies  with known gradients.

\acknowledgments 
We would like to thank the referee for his/her careful reading and constructive comments on the manuscript. 
T.-Y. wants to thank Robert Sharp, Mike Dopita, Peter McGregor and I-Ting Ho for helpful discussions on this work.
T.-Y. acknowledges a Soroptimist Founder Region Fellowship for Women.  
L.K. acknowledges a NSF Early CAREER Award AST 0748559 and an ARC Future Fellowship award FT110101052.  


\end{document}